\documentclass[aps,pra,twocolumn,groupedaddress]{revtex4}

\usepackage{braket}
\usepackage{amsmath}
\usepackage{graphicx}
\usepackage{amssymb}

\begin{document}

%\preprint{}

\title{High-order Harmonic Spectroscopy of the Cooper Minimum in Argon: Experimental and Theoretical Study}


\author{J. Higuet$^1$, H. Ruf$^1$, N. Thir\'e$^2$, R. Cireasa$^2$, E. Constant$^1$, E. Cormier$^1$, D. Descamps$^1$, E. M\'evel$^1$, S. Petit$^1$, B. Pons$^1$, Y. Mairesse$^1$ and B. Fabre$^1$}
%\thanks{}
\affiliation{$^{1}$CELIA, UMR5107 (Universit\'e Bordeaux 1 - CNRS - CEA)\\
 351, cours de la lib\'eration - F-33405 Talence Cedex \textsc{FRANCE}}
\affiliation{$^{2}$Laboratoire Collisions Agr\'egats R\'eactivit\'e, (IRSAMC), UPS, Universit\'e de Toulouse, F-31062 Toulouse, France and CNRS, UMR 5589, F-31062 Toulouse, France}

\date{\today}

\begin{abstract}
We study the Cooper minimum in high harmonic generation from argon atoms using long wavelength laser pulses. We find that the minimum in high harmonic spectra is systematically shifted with respect to total photoionization cross section measurements. We use a semi-classical theoretical approach based on Classical Trajectory Monte Carlo and Quantum Electron Scattering methods (CTMC-QUEST) to model the experiment. Our study reveals that the shift between photoionization and high harmonic emission is due to several effects: the directivity of the recombining electrons and emitted polarization, and the shape of the recolliding electron wavepacket. 
\end{abstract}

\maketitle

\section{Introduction}

High order harmonic generation (HHG) takes place when an atom or a molecule is submitted to a strong laser field 
(with intensities $I$ in the range $10^{13}-10^{14}$ W.cm$^{-2}$). Under the influence of the strong field, an 
electron can be tunnel-ionized, accelerated and driven back to its parent ion where radiative recombination results in 
the emission of XUV (extreme ultraviolet) radiation \cite{Corkum93,Schafer93}. 
While this process was initially considered as a secondary source of radiation, it was soon realized that it is closely 
related to photoionization and that it could thus encode structural information on the irradiated target. The first experimental observation 
that confirmed this link has consisted of the appearance of a local minimum in the high harmonic emission from argon 
atoms \cite{Wahlstrom93}, which was associated to the Cooper minimum observed in XUV photoionization of argon 
\cite{Cooper62}. In photoionization it is well known that this minimum is due to a zero dipole moment between the 
$\textit{p}$ ground state wavefunction and the $\textit{d}$ wavefunction of the photoionized electron for a  
photon energy about 48 eV. Such minima have been extensively studied in photoionization because they constitute 
clearly identifiable features against which theoretical models can be tested. 

The interest for structural minima in HHG has been revived when the case of high order harmonics from aligned molecules 
was considered, showing that minima encoding the molecular structure could appear in harmonic spectra \cite{Lein02}. 
Many works have focused from then on observing these minima  \cite{Itatani04,Kanai05,Vozzi05,Boutu08,Smirnova09b,Worner10,Torres10,Wong10}. 
These studies have raised a number of questions on the modelling of the HHG process such as the influence of the ionic 
potential on the recolliding electron \cite{Le09,Ramakrishna10}, the contribution of multiple molecular orbitals 
\cite{McFarland08,Smirnova09b,Haessler2010}, the influence of the strong laser field \cite{Mairesse10,Sukiasyan10} and the role of propagation effects 
\cite{SickMiller09}. Since some of these questions are still debated, and since HHG from molecules is quite complex to 
model, we decided to come back to the simpler case of high order harmonics from atoms. 

In this paper, we study the Cooper minimum in high order harmonic emission from argon atoms using tunable infrared (1800-2000 nm) femtosecond laser pulses. 
We perform a systematic experimental study of the position of the minimum as a function of the laser field 
(intensity and wavelength) and macroscopic parameters (gas pressure, beam focusing conditions). We find a systematic 
shift of more that 5 eV in the position of the minimum with respect to total photoionization cross section measurements \cite{Marr76,Samson02}. 
We perform a theoretical study to understand the origin of this shift and find that it is partly due to the difference 
between angle-integrated photoionization measurements with unpolarized light and the HHG
recombination process which is highly differential with respect to both electron and polarization directions. 
Using a semi-classical simulation based on a combination of Classical Trajectory Monte Carlo (CTMC) \cite{Abrines66}
and Quantum Electron Scattering techniques (QUEST), we show that the additional contribution to the shift is due to the 
shape of the recolliding electron wavepacket. This new theoretical description, to which we refer to as CTMC-QUEST properly accounts for the influence of the ionic potential on the recolliding electron wavepacket and is 
thus able to reproduce accurately the experimental high order harmonic spectra.

The manuscript is organized as follows. In section II, we describe the experimental setup and present the measurements of 
the Cooper minimum in various laser and macroscopic conditions. In section III, we analyze the analogies and differences 
between photoionization and recombination using quantum electron scattering theory. Finally we perform in section IV 
a complete simulation of the experiment combining CTMC and QUEST approaches. Atomic units are used throughout the paper unless 
otherwise stated.

\section{Experiment}
\subsection{Experimental setup}
It is well known \cite{ Winterfeldt08} that harmonic spectra generally consist of a rapid decrease of the yield for low harmonic orders,
followed by a plateau which extends up to the cut-off photon energy $\sim$3.17$U_p+I_p$, where $U_p$ is the laser ponderomotive
energy $U_p=I/4\omega_0^2$, with $\omega_0$ the laser frequency, and $I_p$ is the target ionization potential
($I_p=$15.76 eV for Ar). Therefore,
an accurate determination of the position of the Cooper minimum in HHG can be achieved provided 
the minimum belongs to the plateau region of the spectrum. Since the minimum lies around 50 eV, this requires using laser 
intensities above $2\times 10^{14}$ W.cm$^{-2}$, close to the saturation intensity of argon when using 800 nm $\sim$ 40 fs laser pulses. A way to 
perform measurements at higher intensities is to use few-cycle laser pulses: W\"orner \textit{et al.} recently measured the Cooper 
minimum in Argon at $I$ up to $3.5 \times 10^{14}$ W.cm$^{-2}$ using 8 fs 780 nm laser pulses \cite{Worner09}. We have chosen an 
alternative solution to obtain a high cut-off: using long wavelength driving laser \cite{Colosimo08,Vozzi09,Shiner2009}. The cutoff frequency 
scales as $I \times \lambda^2$, making it possible to produce 100-eV photons at less than $1\times 10^{14}$ W.cm$^{-2}$ 
using 1800-nm fs pulses. 
\begin{figure}
\begin{center}
\includegraphics[width=.5\textwidth]{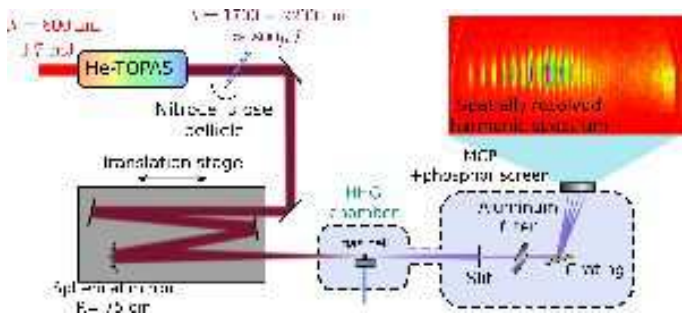}
\end{center}
\caption{\label{FigSetup} Experimental setup (see text).}
\end{figure}

The experimental setup is presented in Fig. \ref{FigSetup}. We use the 1kHz Aurore laser system from CELIA which delivers 
800 nm, 7 mJ, 35 fs pulses. We reduce the pulse energy to 4.7 mJ and inject it to a HE-TOPAS parametric amplifier (Light Conversion Ltd.).
 We use the idler from the TOPAS to get 800 $\mu$J pulses, continuously tunable between 1700 and 2200 nm. 
These pulses are focused by a 75 cm radius spherical silver mirror into a 2 mm 
continuous gas cell filled with argon (backing pressure around 50 mbar). The pulse energy can be adjusted by rotating a 
30 $\mu$m thick nitrocellulose pellicle, whose transmission varies from $\sim 95 \%$ under normal incidence to 
$\sim 50\%$ under $\sim 70^\circ$ incident angle. The high harmonics are sent to an XUV flat field spectrometer consisting of a 250 $\mu$m slit, 
a 1200 mm$^{-1}$ (variable groove spacing) grating (Hitachi), a set of dual microchannel plates associated to a phosphor screen 
(Hamamatsu), and a 12-bit cooled CCD camera (PCO). 

An important item in the experimental setup is a 200 nm-thick aluminum filter placed before the spectrometer. Indeed, high harmonic 
radiation produced by the 1800 nm pulses can extend to more than 100 eV. Our study deals with measurements of the Cooper minimum, 
around 50 eV. Since the high harmonic emission from argon is quite efficient around 100 eV while it is minimum around 50 eV 
\cite{Vozzi09}, the second order diffraction of the 100 eV radiation by the grating can significantly affect the shape of the 
spectrum at 50 eV. We have observed this effect as a bump which fills in the Cooper minimum as the laser energy increases. 
The aluminum foil in our experiment filters out the radiation above 73 eV, enabling us to get rid of this artifact. 

The proper calibration of our XUV spectrometer is crucial to determine
accurately the position of the Cooper minimum. The incidence angle of
the grating is accurately set using a precision rotation mount. The
distances between the source and the grating and between the grating
and MCP are measured. The central wavelength of the fundamental
radiation is measured using a near infrared spectrometer. From all
these parameters we can determine the theoretical position of the
different harmonics on the MCP, assuming a certain order $q_0$ for the
first harmonic. We compare these theoretical positions to the measured
positions and determine the value of $q_0$ which provides good
agreement. Since we measure many harmonics, we can check the fine
agreement of the calculated and measured positions over a broad
spectral range. We can finely adjust the distance value between
the grating and MCP, which is subject to the biggest uncertainty in
our experiment,
 to achieve
perfect agreement and extract the pixel-wavelength conversion function.
%(Fig. \ref{Calibration}(q)). 
As a check of the calibration, we measure the transmission of the
aluminum foil in our experiment. 
%The results are shown in Fig.\ref{Calibration}(b). 
We compare them to the theoretical transmission
of a 200 nm thick filter with a 5 nm layer of alumine on each side
\cite{CXRO}. The position of the abrupt cutoff in the filter
transmission is in excellent agreement.

\begin{figure}
\begin{center}
\includegraphics[width=.45\textwidth]{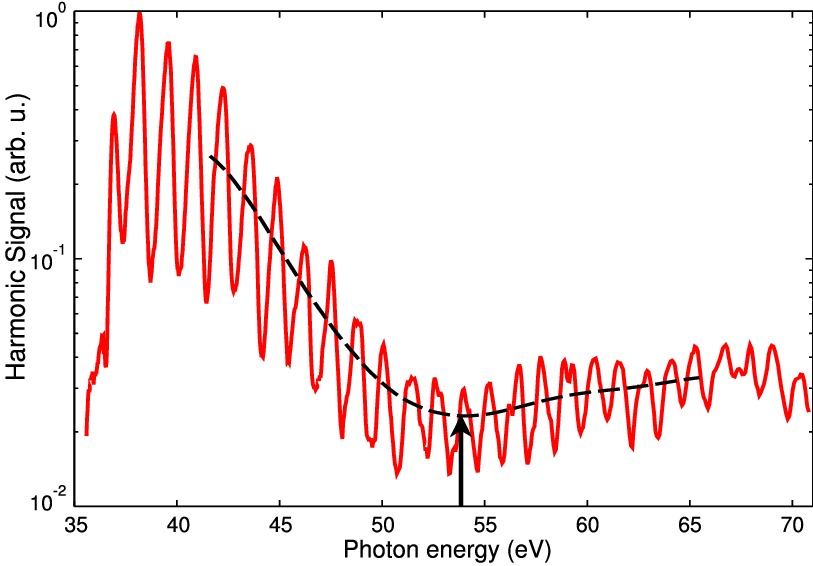}
\end{center}
\caption{\label{FigSpectrum} High harmonic spectrum generated in argon by a 1830 nm $8\times 10^{13}$ W.cm$^{-2}$ laser pulse. The spectrum is averaged over 25000 laser shots. The dashed line is a gaussian smoothing of the spectrum. The arrow indicates the position of the Cooper minimum $E=53.8$ eV.}
\end{figure}

A typical harmonic spectrum obtained at 1830 nm laser wavelength is shown in Fig. \ref{FigSpectrum}. The spectrum was normalized taking into 
account the pixel-wavelength conversion, the diffraction efficiency of the grating and the measured transmission of the aluminum filter. 
An additional advantage of using long wavelength appears in this spectrum: the fundamental photon energy is 0.68 eV so that the 
spectrum is more dense than at 800 nm (to which corresponds a photon energy of 1.55 eV), making the determination of the position 
of the minimum more accurate. 
The dashed line in Fig. \ref{FigSpectrum} is a gaussian smoothing of the  spectrum, from
which we extract the location of the Cooper minimum. We performed a
series of identical measurements to evaluate the average position of the
minimum and obtained $E = 53.8 \pm 0.7 eV$.

\subsection{Systematic study}
In order to check the robustness of the position of the Cooper minimum against experimental conditions, we have performed a 
systematic study of HHG in argon, varying both the laser field parameters, which control the single atom response, and the macroscopic 
parameters (gas pressure, beam focusing conditions), which influence the finally detected HHG yield. 

\subsubsection{Laser field parameters}
An important difference between HHG and photoionization experiments is the presence of a strong laser field, which might modify 
the position of the Cooper minimum. We thus conducted a study as a function of the laser field parameters (intensity and wavelength). 

\begin{figure}
\begin{center}
\includegraphics[width=.45\textwidth]{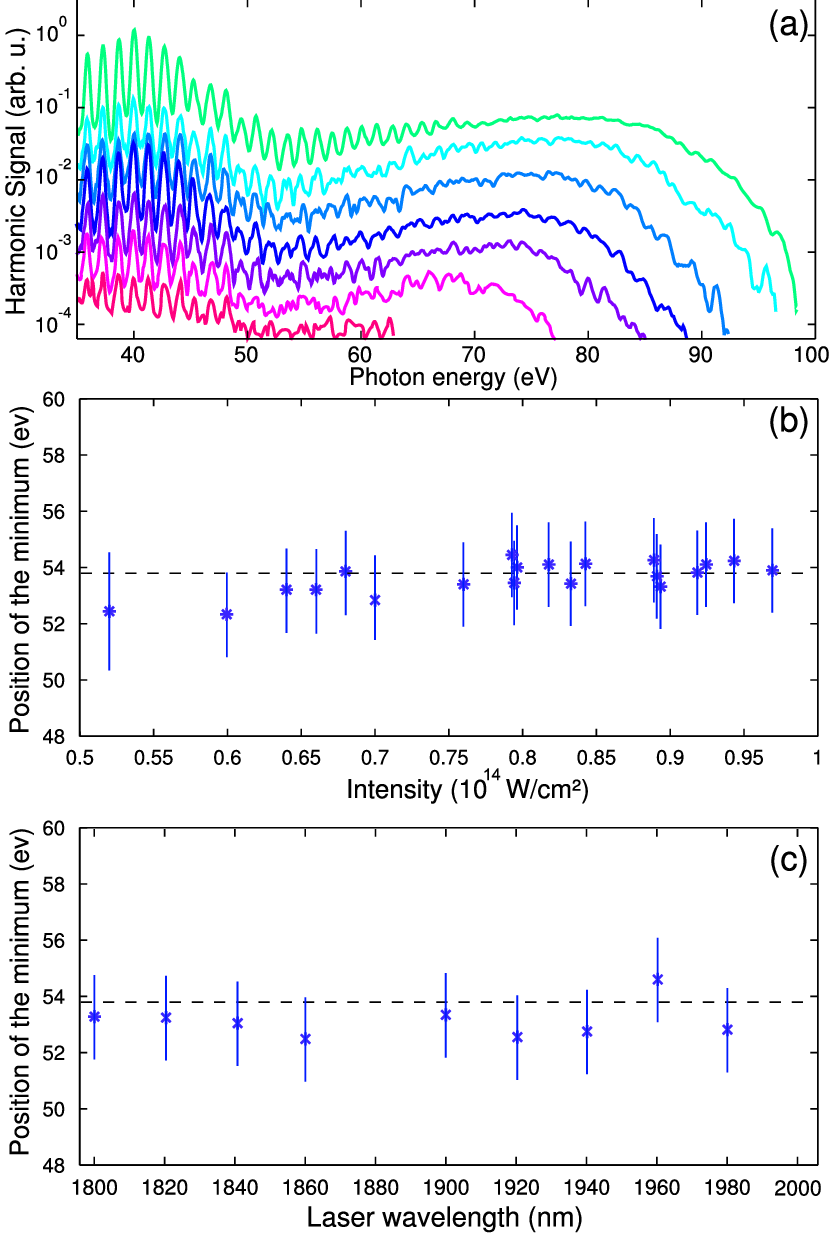}
\end{center}
\caption{\label{FigMinimumVsLaser} (a) High harmonic spectra generated in argon by a 1830 nm laser pulse at intensities between $5\times 10^{13}$ W.cm$^{-2}$ (bottom) and $8\times 10^{13}$ W.cm$^{-2}$ (top). 
No aluminum filter was used for these spectra. The spectra are arbitrarily shifted vertically with respect to each other. 
Each spectrum is averaged over 25000 laser shots. (b) Position of the Cooper minimum in Argon as a function of the laser intensity at 1830 nm and (c) as a function of the laser wavelength. 
The horizontal dashed line marks the minimum location value $E=53.8$ eV determined previously.}
\end{figure}

Figure \ref{FigMinimumVsLaser}(a) shows a few harmonic spectra produced at different laser intensities (controlled by rotating 
the nitrocellulose pellicle). As the intensity increases the cutoff is shifted from 60 eV to 90 eV. The Cooper minimum is clearly 
visible on all spectra. Figure \ref{FigMinimumVsLaser}(b) shows the measured position of the minimum as a function of 
the laser intensity. We do not observe any systematic variation of the minimum location as the intensity is increased twofold even if 
the ponderomotive energy changes by more than 10 eV. 
The positions at lower intensities are slightly below average but the signal being smaller the error bars are more important.
We also varied 
the central wavelength of the laser pulses delivered by the HE-TOPAS between 1800 and 1980 nm (Fig. \ref{FigMinimumVsLaser}(c)). 
We do not observe any significant shift of the Cooper minimum, which stays around 53.8 eV. These two observations constitute strong 
indications of the lack of influence of the laser field on the recombination process. 

\subsubsection{Phase matching}
\begin{figure}
\begin{center}
\includegraphics[width=.45\textwidth]{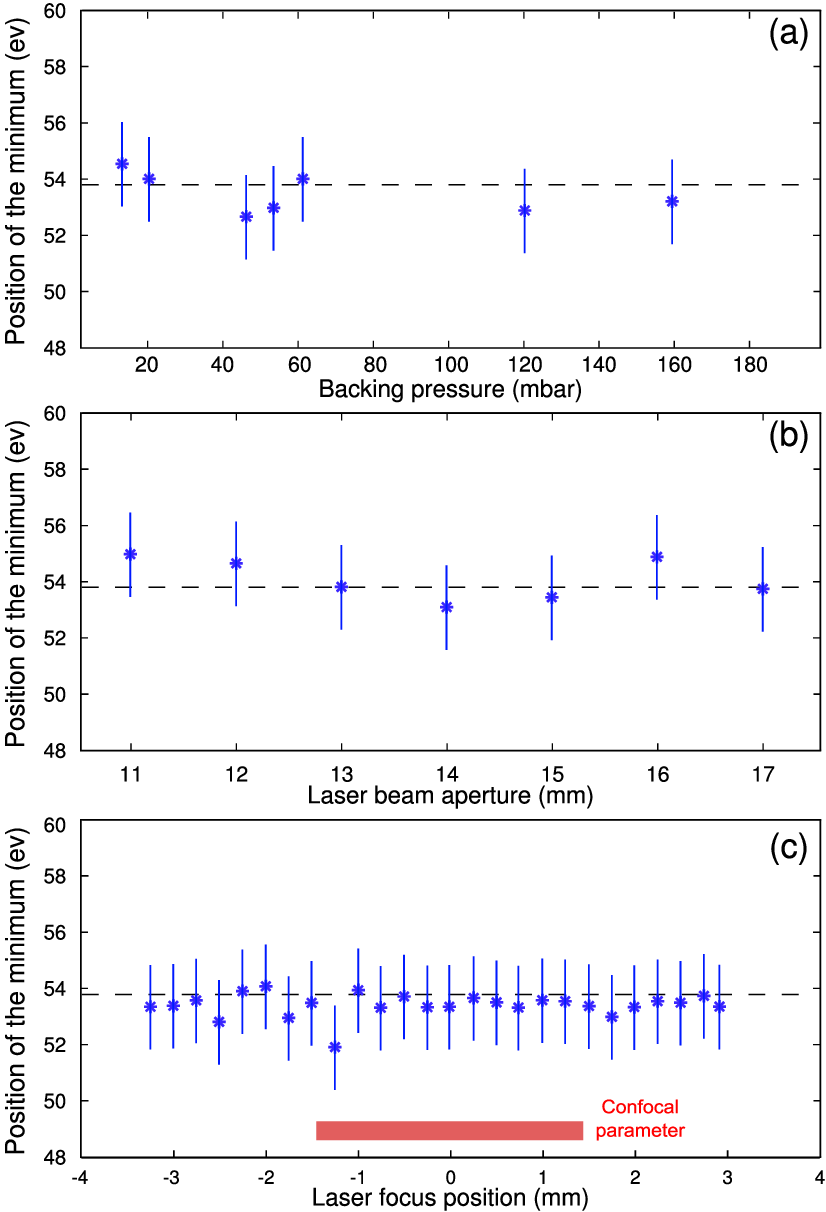}
\end{center}
\caption{\label{FigPhaseMatching} Position of the Cooper minimum in Argon at 1830 nm as a function of the backing pressure in mbar (a)
, the laser beam aperture (b) and longitudinal position of the focus (c). The horizontal dashed line marks the minimum location value $E=53.8$ eV
determined previously.}
\end{figure}

While the Cooper minimum is a characteristic of high order harmonic emission from a single atom, our experiment measures the outcome
of a macroscopic process. It is thus important to evaluate the possible influence of propagation effects in the measured spectra \cite{Salieres95}.
To that purpose, we varied several parameters which affect the phase matching conditions and the macroscopic buildup of the harmonic
signal. First, we varied the backing pressure of the gas cell between 12 mbar and 160 mbar and did not observe any change in the
position of the Cooper minimum (Fig. \ref{FigPhaseMatching}(a)). Second, we changed the laser beam aperture between 11 and 17 mm. This 
resulted in a modification of the harmonic cutoff but no change in the position of the Cooper minimum (Fig. \ref{FigPhaseMatching}(b)). 
Finally, we varied the longitudinal position of the laser focus with respect to the center of the gas cell (Fig. \ref{FigPhaseMatching}(c)). 
This also led to a change of the cutoff position but the value of the minimum location remained the same. 
We note that we did not observe any signature of the presence of long trajectories in our experiment while scanning the focus position: 
the harmonics always appear as spatially collimated and spectrally narrow, which indicates that the short trajectories would always be favored 
in our generating conditions.

These observations are different from what has recently been reported by Farrell \textit{et al.} in a study of the Cooper minimum in Argon using 800 nm pulses \cite{Farrell10}.

\section{Photoionization and radiative recombination}
Our experimental study has shown that the Cooper minimum in HHG is located at 53.8 eV. This position is different from that measured
in photoionization spectra (between 48 and 49 eV). Even though at first sight photoionization and recombination appear as strictly
reverse processes, which would lead to a simple conjugation relation between the associated transition dipole, the experimental
observations show a systematic shift. In this section, we perform a detailed theoretical analysis of the link between photoionization
and radiative recombination processes in order to explain this difference.

Our calculations rely on the Single Active Electron (SAE) approximation \cite{Bransden03} where the description of
electron dynamics is restricted to that of a single Ar valence electron. The interaction of this electron with
the nucleus and other electrons is reproduced, in the framework of a mean-field theory, by the model
potential \cite{Muller99}:
\begin{equation}
V(r)=-\frac{1}{r}-\frac{A\,e^{-Br}+(17-A)\,e^{-Cr}}{r}
\label{potentiel_modele}
\end{equation}
with $A$=5.4, $B$=1, $C$=3.682.
This model potential fulfills  the correct asymptotic condition, {\em i.e.} $ V(r \rightarrow \infty)=-1/r$,
and the expected behavior at the origin, $V(r \rightarrow 0)=-18/r$. In \cite{Muller99}, the parameters $A$, $B$ and $C$
were adjusted so as to provide the correct  $I_p$(Ar); we have further checked that $V(r)$ yields accurate values for the
energies of the Ar($nl$) excited states with $n\ge 4$.

The diagonalization of the hydrogen-like Hamiltonian $\mathcal{H}=p^2/2+V$ in a basis of even-tempered Slater-type orbitals
(STO) yields bound and discretized continuum states in the spherical form
$\Psi_{n,l,m}(\mathbf{r})=R_{n,l}(r)Y^m_l(\Omega)$ where $Y_l^m$ are spherical harmonics of given $(l,m)$ symmetry and
$\Omega_{\mathbf{r}} \equiv (\theta_{\mathbf{r}},\varphi_{\mathbf{r}})$ is the angular
$\mathbf{\hat{r}}$ direction. 
The scattering continuum states $\Psi_\mathbf{k}$, normalized on the wavevector scale $\mathbf{k}$,
are developed on the spherical state basis

\begin{align}
  \Psi_{\mathbf{k}}(\mathbf{r})= \frac{1}{k} \sum^{\infty}_{l=0} \sum^{l}_{m=-l} (\imath)^l e^{\imath\delta_{k,l}}& R_{k,l}(r) \nonumber \\ 
& (-1)^m Y^{m}_{l}(\Omega_\mathbf{r}) Y^{-m}_{l}(\Omega_\mathbf{k})
\label{continuumAr}
\end{align}
where $\Omega_\mathbf{k} \equiv (\theta_\mathbf{k},\varphi_\mathbf{k})$ corresponds to the angular $\mathbf{\hat{k}}$ direction. Since
the diagonalization of $\mathcal{H}$ in the STO underlying basis yields a coarse-grained representation of the continuum \cite{Pons00,Pons00b},
we alternatively obtain the radial parts $R_{k,l}(r)$, and the phase shifts $\delta_{k,l}$ (including both Coulombic and short
range components), through direct integration of the radial Schr\"odinger equation, using the Numerov algorithm.
It is important to note that eq.(\ref{continuumAr}) explicitly includes the influence of the ionic core on the ejected/recombining
electron, beyond the plane wave approximation which does not allow to faithfully describe neither photoionization nor HHG
\cite{Le2008,Worner09}.

\subsection{Photoionization cross section}

The photoionization rate induced by absorption of a radiation of amplitude $F_0$ and polarization axis $\mathbf{n}$ is given by the Fermi's golden rule \cite{Bransden03}:
\begin{align}
 w(\mathbf{k},\mathbf{n})=\frac{\pi}{2} k |\bra{\Psi_{\mathbf{k}}} \mathcal{D} \ket{\Psi_{3,1,0}}|^2
\label{fermi}
\end{align}

where $\mathcal{D}=-F_0 \mathbf{n} \cdot \mathbf{r}$ is the electric dipole moment operator. The photoionization differential cross section is  given by:
\begin{align}
 \frac{d^2\sigma}{d\Omega_{\mathbf{k}}d\Omega_{\mathbf{n}}}=\frac{4\pi^2}{c}k\omega|\bra{\Psi_{\mathbf{k}}} \mathbf{n} \cdot \mathbf{r} \ket{\Psi_{3,1,0}}|^2
\label{copper_sedp}
\end{align}
$\omega$ being the frequency of the ionizing photon, related to the electronic momentum by the relation $\omega=I_p+k^2/2$.

We aim to compare our calculations to photoionization measurements of Marr \& West \cite{Marr76} and Samson \textit{et al} \cite{Samson02}. In these experiments, the measured quantity is the total photoionization cross section, which includes all possible relative orientation between the electronic momentum and light polarization axis:
\begin{align}
 \sigma(k)&=\iint \frac{d^2\sigma}{d\Omega_{\mathbf{k}}d\Omega_{\mathbf{n}}}d\Omega_{\mathbf{k}}d\Omega_{\mathbf{n}}
\label{dipole_tot}
\end{align}

\begin{figure}
\begin{center}
\includegraphics[width=.5\textwidth]{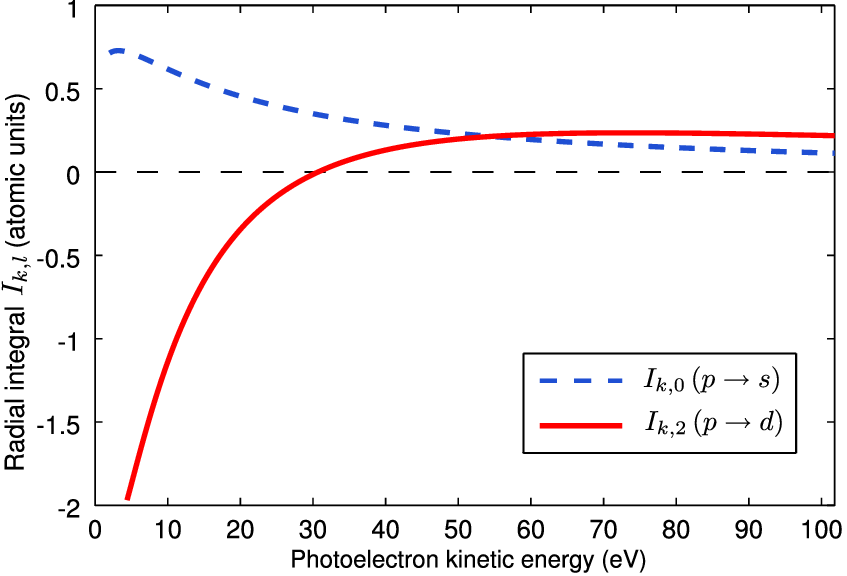}
\end{center}
\caption{\label{FigRadialInt} Radial integrals $I_{k,l}$ associated with $p\rightarrow s$ (dashed line) and $p\rightarrow d$ (plain line) transitions.}
\end{figure}

\begin{figure}
\begin{center}
\includegraphics[width=.5\textwidth]{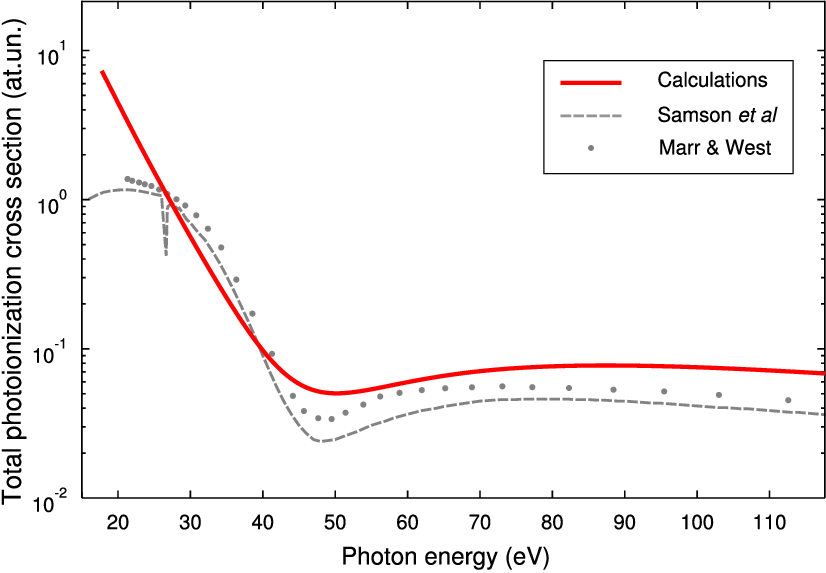}
\end{center}
\caption{\label{FigTotalCS} Calculation of the total photoionization cross section (plain line) compared to measurements from \protect\cite{Marr76} (dotted line) and \protect\cite{Samson02} (dashed line).}
\end{figure}

Because of the selection rules for electric dipole transitions, the calculation of $\bra{\Psi_{\mathbf{k}}} \mathbf{n} \cdot \mathbf{r} \ket{\Psi_{3,1,0}}$ is reduced to the components of $\Psi_{\mathbf{k}}$ whose angular momenta are $l=0$ ($p\rightarrow s$ transition) and $l=2$ ($p\rightarrow d$ transition). The angular part of this calculation can be done analytically, and the total photoionization cross section can finally be expressed as:
\begin{align}
 \sigma(k)=\frac{16\pi^3\omega}{9kc}\left(I_{k,0}^2+2I_{k,2}^2\right)\label{dipole_tot}
\end{align}
where $c$ is the speed of light, and $I_{k,l}$ are the radial integrals $I_{k,l}=\int R_{k,l}(r)R_{3,1}(r)r^3dr$.
These radial integrals are plotted in Fig. \ref{FigRadialInt}. $I_{k,0}$ is positive whatever  $k$ values, and slowly decreases
as the electron energy increases. The $p\rightarrow d$ integral exhibits stronger modulations and a sign change for an electron
kinetic energy of 30.6 eV (corresponding to a photon energy of 46.3 eV). This sign change is responsible for the minimum
observed in the calculated total photoionization cross section, as seen in Fig. \ref{FigTotalCS}.

Our calculation shows a clear minimum in the photoionization cross section for a photon energy of 50.1 eV, slightly above
the measured value (between 48 an 49 eV). The agreement with the experiment is fairly good, except for low photon energies
(below 30 eV), as usual in single active electron calculations. In order to perform accurate calculations in  this energy range,
one has indeed to take into account multielectronic effects and transitions to excited states of the neutral \cite{Parpia1984}.

\subsection{Recombination dipole in HHG}

\subsubsection{Linear laser field}
An important specificity of high order harmonic generation compared to photoionization is the selection of specific
values of $\mathbf{k}$ and $\mathbf{n}$. First, tunnel ionization selects the quantization axis of the atomic orbital ($z$)
parallel to the polarization axis of the IR generating field, as recently confirmed experimentally \cite{Young06,Loh07,Shafir09}.
The IR field drives the trajectory of the recombining electron along the same axis. The electronic momentum $\mathbf{k}$ thus has
to be taken parallel to the quantization axis of the orbital. Furthermore, by symmetry consideration, the emitted XUV field
has to be parallel to the common axis of the laser polarization and quantization. The calculation is thus reduced to the case
of $\mathbf{\hat{k}}\parallel\mathbf{n}\parallel \mathbf{\hat{z}}$.

Within this frame, the recombination matrix element is equal to:
\begin{align}
 |d_{\text{rec}}|^2&=|\bra{\Psi_{\mathbf{k}}} \mathbf{n} \cdot \mathbf{r} \ket{\Psi_{3,1,0}}|^2\nonumber \\
&=\frac{1}{12\pi \,k^2}| I_{k,0}e^{\imath\delta_{k,0}} -2I_{k,2}e^{\imath\delta_{k,2}}|^2
\label{dipole_argonHHG}
\end{align}

\begin{figure}
\begin{center}
\includegraphics[width=.5\textwidth]{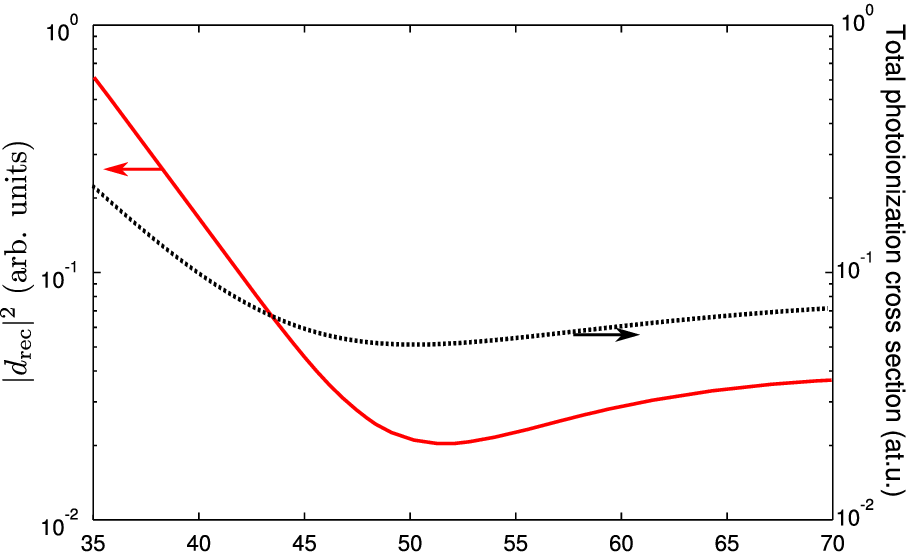}
\end{center}
\caption{\label{FigHarmDipole} Calculation of $|d_{\text{rec}}|^2$ (plain line) compared to total photoionization cross section (dashed line).
}
\end{figure}

The evolution of $|d_{\text{rec}}|^2$ as a function of the emitted photon energy is plotted in Fig. \ref{FigHarmDipole}. The
dipole moment shows a Cooper minimum which is more contrasted than that observed in the total photoionization cross section. This
is due to the coherent nature of the differential cross section: the differential calculation leads to a coherent sum of the
$p\rightarrow s$ and $p\rightarrow d$ transitions, which depends on the values of the scattering phases $\delta_{k,0}$ and
$\delta_{k,2}$. In the case of angular integrated calculation, this phase influence vanishes leading to an incoherent sum of
the two contributions.

The Cooper minimum in the recombination dipole moment is located at 51.6 eV, while it is at 50.1 eV in the total photoionization
cross section. The selection of the quantization axis and light polarization thus lead to a significant shift of the Cooper minimum.
In the following, we confirm the importance of this selection by considering the case of an elliptical laser field.

\subsubsection{Elliptical laser field}
The recollision direction of the electron in HHG can be manipulated using an elliptically polarized laser field. In that case,
the electron trajectory between tunnel ionization and recombination is two-dimensional, the recollision axis is not parallel
to the quantization axis and the polarization direction of the harmonics is not that of the laser field \cite{Antoine97polar}.

We have performed classical calculations to study the influence of ellipticity on the Cooper minimum, in the framework of the
so-called "Simpleman" model \cite{Corkum93} that neglects the influence of the ionic potential on the electron dynamics
between tunnel ionization and recombination. The electron trajectory is supposed to be ionized at time $t_i$ through tunneling
along the instantaneous field direction which sets the quantization axis; this accordingly yields the initial conditions
$\mathbf{r}(t_i)=\mathbf{0}$ and $v_\parallel(t_i)=0$.
We further require the trajectory to be closed, {\em i.e.} $\mathbf{r}(t_r)=\mathbf{0}$ where $t_r$ is the recombination time. This
imposes a non-zero perpendicular velocity at time of ionization, $v_\perp(t_i) \neq 0$, consistently with the lateral confinement
of the electronic wavepacket during tunneling. The recollision angle $\theta_{\mathbf{k}}$, which is defined with respect to the
quantization axis and determines the $\mathbf{\hat{k}}$ direction at time of recombination \cite{Mairesse08ellipt}, is then collected
for each electron trajectory. This finally allows us to compute the parallel $d_\parallel$ and orthogonal $d_\perp$
components of the recombination dipole.

\begin{figure}
\begin{center}
\includegraphics[width=.45\textwidth]{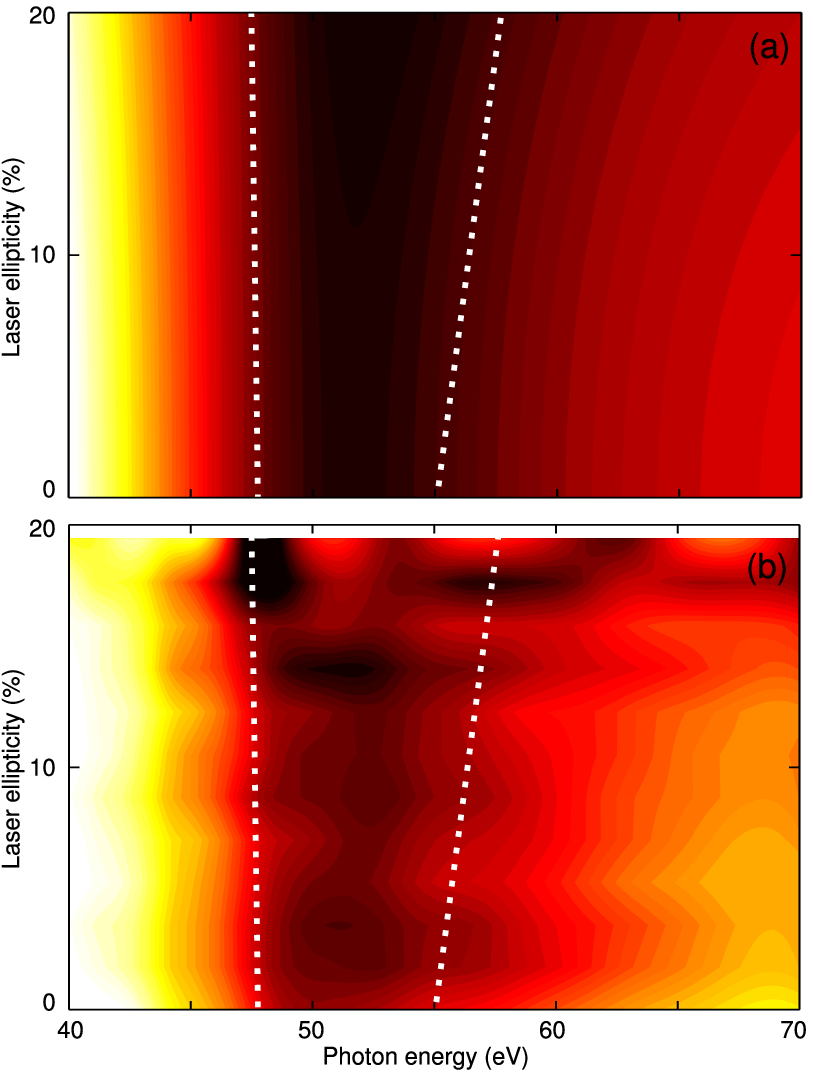}
\end{center}
\caption{\label{FigEllip} (a) Calculation of $|d_{\text{rec}}(k, \theta_{\mathbf{k}}) |^2$ as a function of the photon energy and the ellipticity of the driving field ($\lambda$=1900 nm, I=1$\times$10$^{14}$W.cm$^{-2}$). (b) Measured harmonic spectra in the same conditions. The dotted white lines are the same in the two panels. }
\end{figure}

In Fig. \ref{FigEllip}(a), we represent the value of $|d_{\text{rec}}(k, \theta_{\mathbf{k}})|^2=|d_\parallel+d_\perp|^2$
as a function of the
generating field ellipticity $\epsilon$, for $\lambda=1900$ nm and $I=$1$\times$10$^{14}$W.cm$^{-2}$. As $\epsilon$ increases the overall
shape of the minimum is modified: it becomes broader on the high energy side, with an approximate slope of 0.15 eV by percent
of ellipticity. This behavior stems from increasing contributions of $\theta_{\mathbf{k}} \neq 0$ recombination angles to the HHG signal
as $\epsilon$ increases, given that the minimum of $|d_{\text{rec}}(k, \theta_{\mathbf{k}})|^2$ shifts to higher energy
as $\theta_{\mathbf{k}}$ increases.

We have performed measurements to confirm the broadening of the Cooper minimum. We have added a quarter waveplate in the laser beam.
The ellipticity was controlled by rotating the axis of the waveplate with respect to the laser polarization. As $\epsilon$ increases the
harmonic signal falls rapidly, preventing measurements with ellipticities larger than 20$\%$. 
Figure \ref{FigEllip}(b) shows the normalized harmonic spectra as a function of the laser ellipticity, at an intensity of 
1$\times$10$^{14}$W.cm$^{-2}$ and a laser wavelength $\lambda$=1900 nm. 
The white dotted lines are identical in the two panels, which enables a direct comparison. 
As predicted by our simulations, the Cooper minimum broadens on the high energy side as we increase the $\epsilon$, with an average slope of
approximately 0.2 eV by percent of ellipticity.  
These results confirm the important role of the recollision direction in the accurate determination of the shape of the harmonic spectrum even in a simple atomic system like Ar.

Even though the calculated dipole moments and measured harmonic spectra show very similar shapes, there is still a systematic 2.2 eV
shift between the positions of the measured and calculated Cooper minima. This is due to the fact that the harmonic spectrum
$S(\omega)$ is not only determined by the recombination dipole moment $|d_{\text{rec}}(k, \theta_{\mathbf{k}}) |^2$  but also by the number of
recombining electrons at each energy, {\em i.e.} by the shape of the recolliding electron wavepacket.
In the following, we use a CTMC approach to take this effect into account and compare it to the result of our experiment.

\section{Complete theoretical description: CTMC-QUEST}
\label{REW}
In order to get a full description of the HHG process that includes the influence of the structure of the recolliding electron
wavepacket, we have developed a semi-classical theoretical simulation called CTMC-QUEST (Classical Trajectory Monte Carlo-QUantum
Electron Scattering Theory). Our theory is based on a combination of classical and quantum approaches in the modelling of the
HHG process. It is largely inspired by the well known three-step model of laser-matter interactions \cite{Corkum93}, in which
most of the processes subsequent to primary field-induced ionization are determined by the recollision of the returning wavepacket
on the ionic core. In this respect, it is noteworthy that a Quantitative Rescattering Theory (QRS)  has recently
been developed to successfully describe, e.g., HHG and non-sequential double ionization processes based on the
recollision picture \cite{Le09,Micheau09}. In the QRS, the returning electron wavepacket is obtained by means of the strong field approximation
(SFA) \cite{Lewenstein94} or time-dependent Schr\"odinger calculations of an easily solvable model system with similar $I_p$ than that of
the target system effectively considered. Our semi-classical description aims at avoiding such a requirement of
approximated or additional computations; we therefore use CTMC simulations to build the returning wavepacket in terms
of electron trajectories which fulfill preselected rescattering conditions. Once the wavepacket is defined, quantum
rescattering theory is used to compute the recombination probability.

\subsection{CTMC and initial phase-space distribution}
The CTMC method has originally been developed to describe non-adiabatic processes in atomic collisions \cite{Abrines66}. It has
been successfully applied to shed light on subtle ionization mechanisms \cite{Stolterfoht97,Shah03} that cannot be represented unambiguously by purely quantum mechanical
treatments. CTMC calculations have also been performed to describe laser-matter interactions
see e.g. \cite{Botheron09,Cohen01,Uzdin10}. As in atomic collisions, the classical assumption inherent in the CTMC method is not
prohibitive in the description of field-induced processes; for instance, CTMC trajectories leading to HHG have been shown
to match their quantum counterparts obtained in the framework of an hydrodynamical Bohmian description of the electron flow
\cite{Botheron10}.

The CTMC approach employs a $\mathcal{N}$-point discrete representation of the phase-space distribution $\varrho(\mathbf{r},\mathbf{p},t)$
in terms of independent electron trajectories $\{ \mathbf{r}_j(t), \mathbf{p}_j(t) \}$
\begin{align}
\varrho(\mathbf{r},\mathbf{p},t)=\frac{1}{\mathcal{N}} \sum_{j=1}^{\mathcal{N}}{\delta(\mathbf{r}-\mathbf{r}_j(t))
\delta(\mathbf{p}-\mathbf{p}_j(t))}
\label{rho_C}
\end{align}
where $\mathbf{p}$ is the canonical momentum conjugate to $\mathbf{r}$. The temporal evolution of the $\varrho$ distribution is
governed by the Liouville equation, which is the classical analogue to the time-dependent Schr\"odinger equation

\begin{align}
\frac{\partial \varrho(\mathbf{r},\mathbf{p},t)}{\partial t} =-\frac{\partial \varrho(\mathbf{r},\mathbf{p},t)}{\partial \mathbf{r}}
\frac{\partial H}{\partial\mathbf{p}} +
\frac{\partial \varrho(\mathbf{r},\mathbf{p},t)}{\partial \mathbf{p}}\frac{\partial H}{\partial \mathbf{r}}
\label{liouville}
\end{align}
where $H=\mathcal{H}+\mathbf{r} \cdot \mathbf{F}(t)=p^2/2+V(r)+\mathbf{r} \cdot \mathbf{F}(t)$ is the total electronic Hamiltonian, including the laser-target
interaction $\mathbf{r} \cdot \mathbf{F}(t)$ expressed in the length gauge. 
In our calculations, we use a simple sinusoidal laser field
$F(t)=F_0 \sin(\omega t)$ where $t \in [ 0,4\pi/\omega ]$. This description reduces the interaction of a real shaped
laser pulse with a macroscopic medium to a simplified scheme where the harmonic signal is generated by a single atom
submitted to an effective laser field intensity. This simplification is justified by the insensitivity of the
experimental results to macroscopic effects and to the laser intensity.

Inserting eq.(\ref{rho_C}) into eq.(\ref{liouville}) yields the Hamilton equations which tailor the motion of
the $j^{th}$ classical trajectory among the set of $\mathcal{N}$ independent ones
\begin{align}
\left\{ \begin{array}{lll}
     \frac{\partial \mathbf{r}_j(t)}{\partial t} & = & \mathbf{p}_j(t)   \\
    \frac{\partial \mathbf{p}_j(t)}{\partial t} & = & - \nabla_\mathbf{r}( V(r) +\mathbf{r} \cdot \mathbf{F}(t) ) |_{\mathbf{r}_j(t)}
   \end{array}
\right.
\end{align}

%\begin{eqnarray}
%     \frac{\partial {\bf r}_j(t)}{\partial t}  =  {\bf p}_j(t)  \\
%    \frac{\partial {\bf p}_j(t)}{\partial t}  =  - \nabla_{\bf r} ( V(r) +{\bf r.F(t)} ) |_{{\bf r}_j(t)} \nonumber
%\end{eqnarray}

The last equation emphasizes how the electron evolves under the combined action of the ionic and laser fields,
beyond the widely used SFA that neglects $V(r)$; this has important consequences on the description of radiation
and electron emission processes close to the ionization threshold \cite{Smirnova08,Soifer10}.

\begin{figure}
\begin{center}
\includegraphics[width=.5\textwidth]{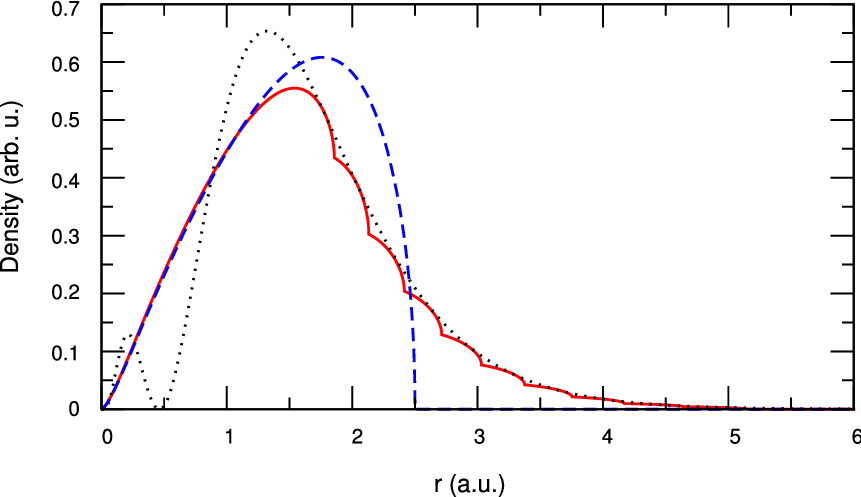}
\end{center}
\caption{\label{FigInitDistrib} Comparison between the initial density  $\rho(r,t=0)$ as a linear combination of 10-microcanonical distribution (plain line), $\rho_M(-I_p;r,t=0)$ the one-microcanonical distribution (dashed line) and the quantum radial electronic density of the fundamental Ar $3p$ state (dotted line).}
\end{figure}

The most common way to define an initial distribution is to use a microcanonical distribution whose energy is $-I_p$.
In Fig. \ref{FigInitDistrib}, we compare the quantum radial probability density $4\pi r^2R^2_{3,1}(r)$ of the initial $3p$ Ar state
to its microcanonical counterpart $\rho_M(-I_p;r,t=0) =\iint{\varrho_m(-I_p;\mathbf{r},\mathbf{p},t=0)d\mathbf{p}d\Omega_{\mathbf{r}}}
 = 16 \pi^2 r^2 \mathcal{C} \sqrt{2(-I_p+V(r))}$ where $\mathcal{C}$ is a normalization constant.
Besides the fact that $\rho_M(r,t=0)$ does not reproduce the nodal structure of the quantum $R^2_{3,1}(r)$,
it restricts the electron density to an inner region close to the nucleus ($r \lesssim 2.5$) where ionization processes
are classically less probable. Such distribution can thus  not properly describe the ionization.

Better quantitative results can be obtained in the framework of the CTMC approach provided that
one employs an improved initial distribution $\varrho(\mathbf{r},\mathbf{p},t=0)$, beyond the usual microcanonical
description $\varrho_m(-I_p;\mathbf{r},\mathbf{p},t=0)$ which affects $-I_p$ to the energy of all $\mathcal{N}$ trajectories \cite{Hardie83}.

Improving the initial condition is based on the observation that a trajectory
with energy $E_j(t=0)=p_j^2(t=0)+V(r_j(t=0)) \neq -I_p$ can participate to the description of the initial state since the
negative part of the classical energy scale is not quantized. In other words, one can construct an improved initial phase-space
distribution as a functional integral over microcanonical ones
\begin{align}
\varrho(\mathbf{r},\mathbf{p},t=0) =\int_{E^-}^{E^+}{c(E)\varrho_M(E;\mathbf{r},\mathbf{p},t=0)dE}
\label{improved}
\end{align}
where the bounds $E^-$ and $E^+$ are given by the partition of the classical phase space into adjacent and non-overlapping
energy bins $[ E^-,E^+[$ for given $l$. The partition is made following the method explained in \cite{Rakovic01}; for Ar($3p$)
this leads to $E^-=-1.885$ and $E^+=-0.150$ for a classical momentum $|\mathbf{L}|=|\mathbf{r} \times \mathbf{p}|$ enclosed in $[1,2[$.
In practice, the improved $\varrho(\mathbf{r},\mathbf{p},t=0)$ consists of a discrete representation of eq.(\ref{improved}) in terms
of 10 microcanonical distributions; the (discretized) coefficients $c_{E_m}$ are obtained by fitting the
quantum radial probability density to
$\rho(r,t=0)=\sum_{m=1}^{10}{c_{E_m}\rho_M(E_m;r,t=0)}$ with the additional condition that $<E>_\rho=-I_p$. 

The final $\rho(r,t=0)$ is displayed in
Fig. \ref{FigInitDistrib}; while the nodal structure of the quantum density still remains beyond the scope of the improved classical description,
as expected, the newly defined $\rho(r,t=0)$ nicely matches the outer region of $R^2_{3,1}(r)$ from which classical electrons
preferentially escape subject to the laser field. Finally, our improved calculations employ $\mathcal{N}=20 \times 10^6$
trajectories in order to minimize statistical uncertainties.

\subsection{CTMC and returning electrons}

\begin{figure}
\begin{center}
\includegraphics[angle=-90,width=.5\textwidth]{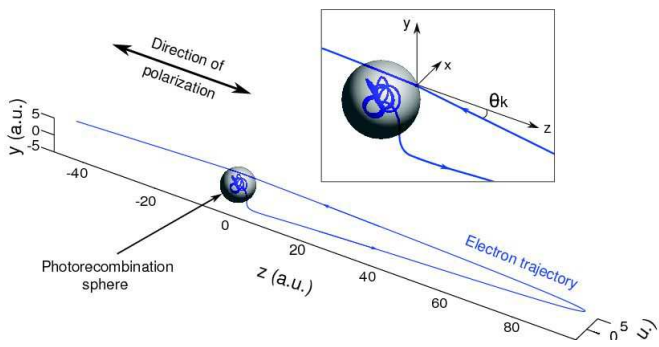}
\end{center}
\caption{\label{FigTrajectoire} Typical short electron trajectory that contributes to the harmonic signal. The electron after being ionized and leaving the 
sphere comes back into it. The inset shows the recombinaison angle $\theta_k$ when the electron is recombining. }
\end{figure}

We now address the definition of the rescattering electron wavepacket.
We define a rescattering sphere, centered on the target nucleus, of radius $R_{\text{rec}}$ of the order of the extension of the
fundamental wavefunction. We consider that an electron is rescattering, and will generate an harmonic photon through
recombination, if after being ionized and leaving the sphere, it comes back into it. The Fig. \ref{FigTrajectoire} shows a typical 
short electron trajectory which contributes to the harmonic signal. The recombination time is taken as
the time the electron enters back into the sphere. We record the energy $E_j$ and the direction $\mathbf{\hat{k}}$
of the wavevector at this instant. Within the CTMC statistics, we are thus finally able to define the density
of the returning wavepacket at time $t$
\begin{align}
\varrho_{ret}(E,\mathbf{\hat{k}},t)=\frac{1}{\mathcal{N}}\sum_{i=1}^{\mathcal{N}_{ret}(t)}{\delta (E_i(t)-E) \delta (\mathbf{\hat{k}}_i(t)-
\mathbf{\hat{k}}})
\label{ret}
\end{align}
where ${\mathcal{N}_{ret}(t)}$ is the number of electron trajectories that fulfill at time $t$ the rescattering criterion
mentioned above.

\begin{figure}
\begin{center}
\includegraphics[width=.5\textwidth]{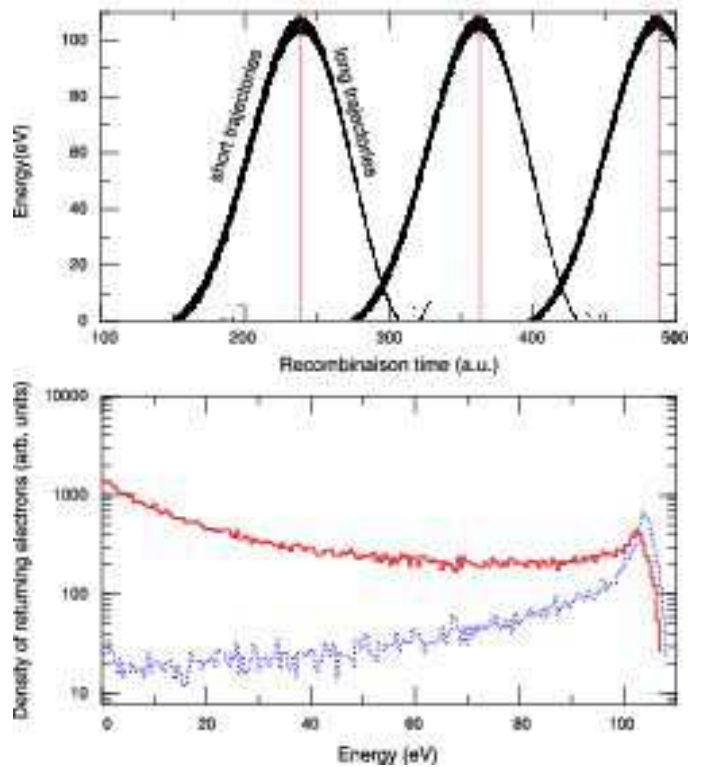}
\end{center}
\caption{\label{FigLongVsShort} (a) : Total energy of the returning electron as function of the photorecombination time. This representation permits a direct and intuitive separation of the different trajectories. In this figure, we only represent the long and short trajectories and omit the multiple returnings.
(b) : Representation of the density of returning electrons for each type of trajectory. }
\end{figure}

Figure \ref{FigLongVsShort} shows a typical result for $I=10^{14}$ W.cm$^{-2}$ and $\lambda=1830$ nm. By monitoring the energy of the
returning electron with respect to the recombination time (Fig. \ref{FigLongVsShort}(a)) we can identify the short (plain line) and long (dotted line)
trajectories. This is an important asset of CTMC-QUEST: the separation of the different classes of trajectories
(short, long, or even multiple return ones) is natural.

In Fig. \ref{FigLongVsShort}(b), we plot the time- and angle-integrated density of returning electrons for long and short trajectories
\begin{align}
\rho_{ret}(E)=\int_0^{\tau}{\int{\varrho_{ret}(E,\mathbf{\hat{k}},t) d\mathbf{\hat{k}}dt}}
\end{align}

The density of returning electrons $\rho_{ret}(E)$ increases with energy for long
trajectories whereas it decreases for short ones. In other words, $\rho_{ret}(E)$ decreases as the
trajectory length increases. Two opposite effects are at play to determine the evolution of $\rho_{ret}(E)$ with trajectory
length. First, longer trajectories are born at earlier times when the laser field is more intense and thus benefit from a
stronger ionization rate. However longer trajectories also show a stronger lateral spread during propagation in the continuum,
which reduces the number of recombining electrons. Our study shows that the latter effect is dominant. It is consistent with the fact that we do not observed long trajectories in our experimental study.

\begin{figure}
\begin{center}
\includegraphics[width=.5\textwidth]{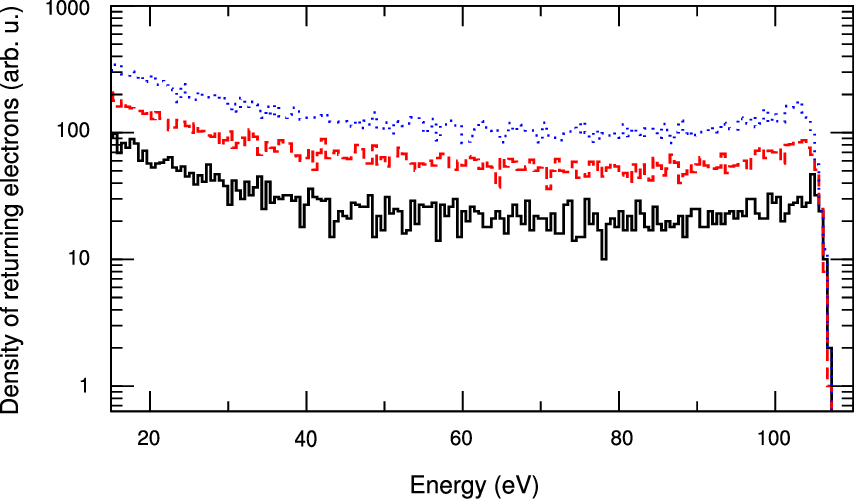}
\end{center}
\caption{\label{FigConvergence} Comparison of the convergence for different radius size $R_\text{rec}=3$ (plain line), 5 (dashed line) and 7 a.u. (dotted line) of the photorecombination sphere for the short trajectories.}
\end{figure}

In Fig. \ref{FigConvergence} we present $\rho_{ret}(E)$ for several values of the photorecombination sphere radius $R_{\text{rec}}$.
This figure enables us to check the convergence of our calculation. In the plateau region, there is no significant difference between the 
shapes of the wavepackets issued from the 3 calculations, which means that the convergence is good. In the following, we will use a photorecombination sphere of $R_{\text{rec}}=5$ a.u., which roughly corresponds to the extension of the fundamental
Ar wavefunction (see Fig. \ref{FigInitDistrib}).

Our results can be compared to those obtained by Levesque \textit{et al.} \cite{Levesque07atoms} who used experimental spectra and dipole moments
calculated within the plane wave approximation (PWA) to extract the amplitude of the recolliding electron wavepacket. 
The lower wavelength of the driving field (800 nm) employed by Levesque \textit{et al.} cannot
explain the sharp disagreement between their and our results: while the shape of our wavepacket falls down by (only) a factor
of $\sim 5$ in a $100$ eV energy interval, several orders of magnitude appear between the low and high energy signals in
\cite{Levesque07atoms}. 
We checked that this discrepancy is due  to the use of  PWA by dividing our experimental HHG spectrum of Fig. \ref{FigSpectrum} by the square of the dipole moment
computed in the PWA and recovering the drastic decrease of $\rho_{ret}(E)$ as $E$ increases.
This confirms the importance of taking into account the influence of the ionic potential in all the three steps of the laser-matter interaction.

%In fact, we have traced the root of this discrepancy back to the use of the PWA; if we proceed
%as Levesque \textit{et al.} and , . Therefore, such a behavior
%stems from an inadequate recovering of $\rho_{ret}(E)$ and reiterates the fact that one has to make particular allowance
%for 
We can also compare our results to the Quantitative Rescattering Theory (QRS) calculations from Le \textit{et al.} (see Fig.2 of \cite{Le09}). In the QRS
framework, the amplitude of the electron wavepacket slightly depends on the method employed to compute the HHG spectrum.
Le \textit{et al.} display a flat-shaped wavepacket amplitude, as a function of $E$, when TDSE calculations are used to
compute the spectrum before dividing it by the dipole moment. Significant deviations are obtained at low $E$ when the SFA
is employed. We have performed (but do not show for sake of conciseness) similar calculations to those of Le \textit{et al.}
who employed a 800 nm driving pulse. This has allowed us to ascertain that at 800 nm the shape of $\rho_{ret}(E)$ is
indeed flat from low to high $E$ when integrating for all trajectories and multiple returns as it is the case in TDSE calculations.
% In fact, $\lambda$ controls, at given $I$, the ponderomotive energy $U_p$ which in turn
%strongly influences the lateral spread of the electron wavepacket between ionization and recollision; the longer $\lambda$,
%the larger is the lateral spread of the wavepacket which induces the falling $E$-shape of $\rho_{ret}(E)$.

\subsection{QUEST}

The photorecombination probability $\mathcal{P}_{PR}(E,\mathbf{\hat{k}},t)$ is quantum mechanically obtained by using the microreversibility
principle and the Fermi's golden rule :
\begin{align}
 \mathcal{P}_{PR}(E,\mathbf{\hat{k}},t)&= \mathcal{P}_{PI}(\mathbf{k'};t) \nonumber \\
&=\frac{\pi \sqrt{2E}}{2} |\bra{\Psi_{E,\mathbf{\hat{k}}}} \mathbf{n} \cdot \mathbf{r} \ket{\Psi_{3,1,0}}|^2
\end{align}
where $\Psi_{E,\mathbf{\hat{k}}}$ is the stationary scattering wavefunction for a total energy $E$ and a wave vector direction $\mathbf{\hat{k}}$.
In the case of photoionization, the electron wavevector $\mathbf{k'}$ is related to the kinetic energy at infinity
through $E= k'^2/2$. Recombination occurs at short range where the total electron energy reads $E= k^2/2+V(R_{\text{rec}})$.
The analogy between photoionization and photorecombination is thus only respected if one sets $k'=\sqrt{2E}$.

In the first part of the manuscript, we calculated photorecombination probabilities assuming that the electron trajectory was
linear and parallel to the quantization axis. In the present statistical 3D modelling, this is no more the case: the electron
escapes from $\mathbf{r} \neq \mathbf{0}$ with a non-zero transverse velocity; its interaction with the ionic potential can further
deviate its trajectory from a straight line. Therefore, it is necessary to take  the recombination angle
$\theta_k$ into account, defined with respect to the $z$-axis, in the computation of the photorecombination probability
\begin{align}
\mathcal{P}_{PR}(E,\mathbf{\hat{k}},t)=\frac{\sqrt{2E}}{24} &  |I_{E, \mathbf{\hat{k}},0}e^{i \delta_{E,\mathbf{\hat{k}},0}} \nonumber \\
& -I_{E,\mathbf{\hat{k}},2}\left (3\cos^2{\theta_k}-1 \right )e^{i \delta_{E,\mathbf{\hat{k}},2}}  |^2
\end{align}

\subsection{CTMC-QUEST for HHG}
Once the density of returning electrons and photorecombination probabilities are determined, we can calculate the
harmonic spectrum $S(\omega,\mathbf{\hat{n}},t)$ as:

\begin{align}
S(\omega,\mathbf{\hat{n}},t) & =  \int{dE \varrho_{rec}(E,\mathbf{\hat{k}},t)\mathcal{P}_{PR}(E,\mathbf{\hat{k}})\delta(E+I_p-\omega)} \nonumber \\
& = \frac{1}{\mathcal{N}}\sum_{i=1}^{\mathcal{N}_{ret}(t)}  {\mathcal{P}_{PR}(E_i(t),\mathbf{\hat{k}}_i(t))f(E_i(t),\omega)}
\label{spectre}
\end{align}

where $f(E_i(t),\omega)=1$ if $E_i(t)=\omega-I_p$ and $0$ otherwise. In case of a laser field linearly polarized along the $z$-axis $\mathbf{\hat{n}}=\mathbf{\hat{z}}$ . The time-integrated
HHG spectrum is simply obtained through $S(\omega,\mathbf{\hat{n}},\tau)=\int_0^{\tau}{S(\omega,\mathbf{\hat{n}},t)}$ where $\tau$ is the pulse
duration.
%In practice, ${\mathcal{P}_{PR}(E_i(t),{\bf \hat{k}}_i(t))}$ consists of a statistical distribution
%in the $(E,t)$ plane that we convert into a spectrum by applying a combination of energy binning and kernel
%gaussian smoothing.

\subsubsection{Role of the returning electron wavepacket}
\begin{figure}
\begin{center}
\includegraphics[width=.5\textwidth]{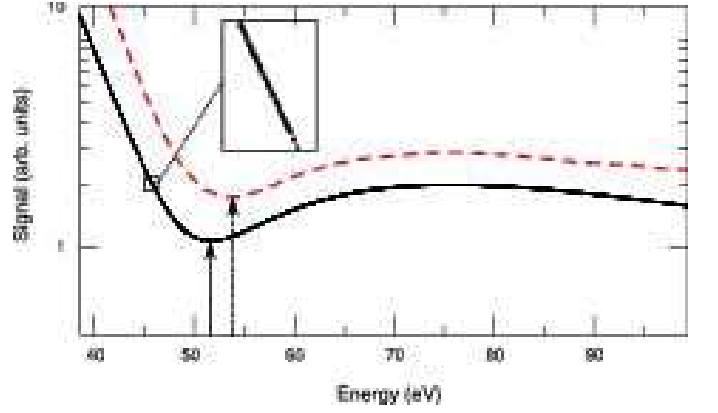}
\end{center}
\caption{\label{FigHarmVsDip} Comparison between the statistical distribution of  recombination probabilities (scattered points which look like a plain line, see inset for details) and the harmonic spectrum (dashed line) for short trajectories obtained by CTMC-QUEST. The position of the Cooper minimum is shifted towards high energy for the harmonic spectrum. This is a consequence of the shape of the density of returning electrons.}
\end{figure}

In Fig. \ref{FigHarmVsDip}, we display the statistical distribution of the recombination probabilities
$\mathcal{P}_{PR}(E_i(t),\mathbf{\hat{k}}_i(t))$ associated to all the short trajectories that lead to HHG during the whole
interaction. This distribution consists of a set of scattered points in the $(E,\mathcal{P}_{PR}(E,\mathbf{\hat{k}}))$ plane
(see the inset in Fig. \ref{FigHarmVsDip}). 
The distribution is very narrow since $\theta_{k_i} \sim 0$ for most
of the trajectories so that its shape is similar to that of the dipole of Fig. \ref{FigHarmDipole}.

The time-integrated
HHG spectrum $S(\omega,\mathbf{\hat{n}},\tau)$ built according to eq.(\ref{spectre}) with only short trajectories is
included in Fig. \ref{FigHarmVsDip}.
The position of the Cooper minimum in $S(\omega,\mathbf{\hat{n}},\tau)$, located at $E=53.5 \pm 0.5$ eV, is different from the one
obtained in the probability distribution. The latter one, found at $E=51.7 \pm 0.5$ eV, is close to the value obtained
assuming $\theta_{k_i}=0$ for all the trajectories. The difference of minimum locations is mainly due to the shape of the recolliding
electron wavepacket for short trajectories (Fig. \ref{FigLongVsShort}(b)): the decrease of $\rho_{ret}(E)$ with increasing $E$
shifts the minimum observed in the photorecombination probability distribution to higher energies. This result, obtained on a
simple atomic system, demonstrates that the accurate study of orbital structure via HHG cannot be restricted to the study of
the dipole moment. The shape of the returning wavepacket has also to be accurately known.

\begin{figure}
\begin{center}
\includegraphics[width=.5\textwidth]{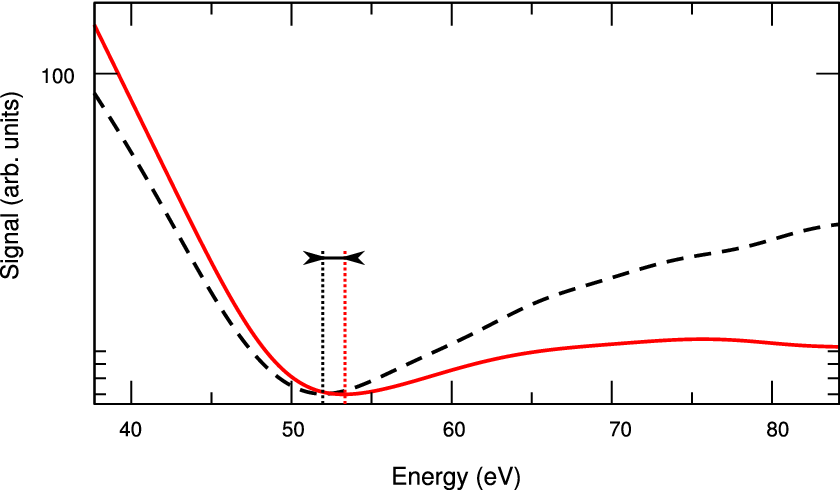}
\end{center}
\caption{\label{FigShortLongHarm} Comparison between the theoretical harmonic spectrum obtained either for long (dashed line) and short (plain line) trajectories. 
As expected looking to the shape of the returning wavepacket, the position of the Cooper minimum is different for these two types of trajectories.}
\end{figure}

The importance of the shape of the recolliding wavepacket is confirmed in Fig. \ref{FigShortLongHarm} where we compare the harmonic
spectra generated by the short and the long trajectories. Interestingly, the position of the Cooper minimum is sensitive to
the considered set of trajectories: it stands at 51.7 eV for long trajectories and at 53.7 eV for short ones. This behavior is
due to the opposite slopes of the electron wavepacket profiles (see Fig. \ref{FigLongVsShort}(b)).

\subsubsection{Comparison with experiments}
 \begin{figure}
\begin{center}
\includegraphics[width=.5\textwidth]{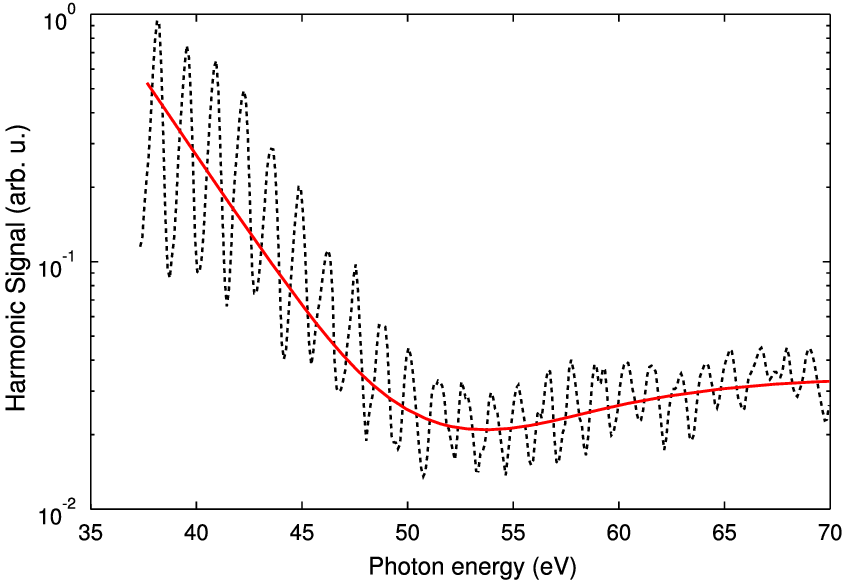}
\end{center}
\caption{\label{FigExpVsTh} Comparison between an experimental (dashed line) and CTMC-QUEST (full line) spectra obtained for with a 1830 nm laser field at $1\times10^{14}$W.cm$^{-2}$.}
\end{figure}

Figure \ref{FigExpVsTh} shows a comparison between the experimental and theoretical HHG spectra for $\lambda=1830$ nm and
$I=10^{14}$ W.cm$^{-2}$. The agreement is very satisfactory since
%even though the CTMC-QUEST approach cannot reproduce
%the observed peaks because of the classical (i.e. continuous) definition of $\varrho_{ret}(E,\mathbf{\hat{k}},t)$. 
the position of the theoretical Cooper minimum is 53.7 eV which nicely matches the experimental one, 53.8 $\pm$ 0.7 eV.
The overall shape of the experimental spectrum is also very well described.

\begin{figure}
\begin{center}
\includegraphics[width=.5\textwidth]{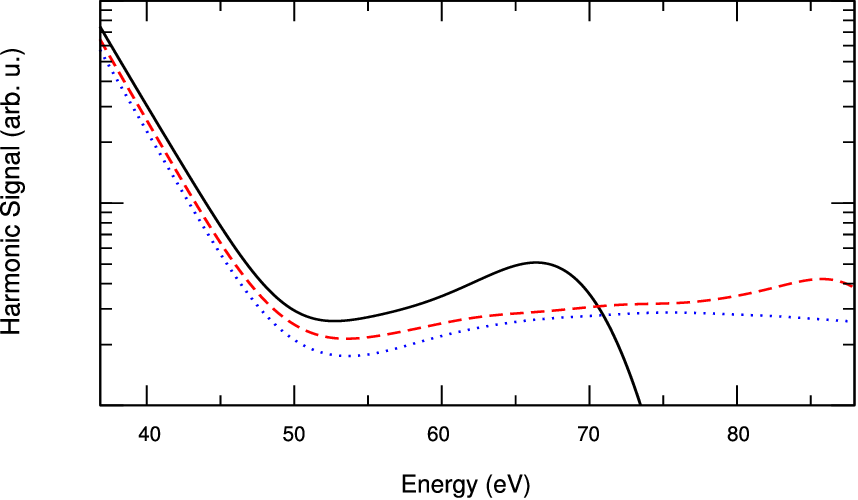}
\end{center}
\caption{\label{FigIntensity} CTMC-QUEST spectra obtained with a 1830 nm laser field at $5\times10^{13}$ (plain line), $7.5\times10^{13}$ (dashed line) and $1\times10^{14}$ W.cm$^{-2}$ (dotted line).}
\end{figure}

In Fig. \ref{FigIntensity}, we compare the results of our CTMC-QUEST calculations for three different laser intensities $I=5$, 7.5 and 10
$\times10^{13}$ W.cm$^{-2}$. There is no clear variation of the position of the Cooper minimum between the two highest intensities.
However, at the lowest intensity the minimum is shifted to lower energies (52.7 eV). This reflects the intensity dependence of the
shape of the rescattering wavepacket. Our experiments are consistent with the lack of variation of the minimum
between $7.5\times10^{13}$ and $1.0\times10^{14}$ W.cm$^{-2}$. At lower intensities we observed a slight downshift of the minimum
which could be the signature of the modification of the wavepacket described by the theoretical results.

\section{Conclusion}
In this paper we have studied the link between photoionization and high harmonic generation by focusing on the Cooper
minimum in Argon. This structural feature is common to the two processes but while the position of the minimum
is observed between 48 and 49 eV in total photoionization cross sections, we measured it at 53.8 eV in HHG. By performing
a systematic experimental study using 1800 nm laser pulse, we have concluded that the position of the Cooper minimum was
independent on the gas pressure and focusing conditions and to a large extent to the laser intensity.

The observed shift of the Cooper minimum is partly due to the fact that the recombination process involved in HHG is the
inverse process of a particular case of photoionization, in which the quantization axis of the atomic orbital, electron ejection
direction and XUV photon polarization are parallel. We have checked the influence of these parameters by manipulating the electron
trajectory using ellipticity, and have observed similar effects in recombination dipole moments and experimental harmonic spectra.

The dipole moment is not sufficient to determine accurately the harmonic spectrum: an additional shift of the Cooper minimum
is due to the structure of the recombining electron wavepacket.

In order to reproduce the experimental observations, we have developed a model based on the combination of CTMC and
Quantum Electron Scattering techniques. CTMC-QUEST takes into account the role of the ionic potential on the
3D trajectories of the electrons in the continuum. The recombination dipole moment corresponding to each
individual electron trajectory is quantum mechanically computed taking into account the recollision angle with respect to the quantization axis. 
This
procedure enables us to obtain a very satisfactory agreement between theoretical and experimental spectra.
It further allows to unambiguously discriminate between short, long and multiple return trajectories contributing to the recollision wavepacket.

Our results show that accurate high harmonic spectroscopy can be performed using long wavelength lasers. The degree of precision
reached by the experiment has allowed us to refine our theoretical modelling. We are now working on the extension of CTMC-QUEST
to polyatomic molecules  in the perspective of achieving complete 3D description of the generation process \cite{Le09}.

\begin{acknowledgments}
We acknowledge financial support from the ANR (ANR-08-JCJC-0029 HarMoDyn) and from the Conseil Regional d'Aquitaine (20091304003 ATTOMOL and COLA project).
We also thank  Val\'erie Blanchet, Nirit Dudovich, Dror Shafir and Andrew Shiner for fruitfull discussions. 
\end{acknowledgments}

\bibliography{biblio.bib}

\end{document}